\documentclass[jmp, showpacs, showkeys, twocolumn]{revtex4-1}

\usepackage{amsmath}
\usepackage{bbold}
\usepackage{graphicx}
\usepackage{tikz}
\usetikzlibrary{positioning}	

\newcommand{\rep}[1]{{\bf #1}}
\newcommand{\repbar}[1]{{\bf \overline{#1}}}

\begin{document}

\title{Braids as a representation space of SU(5)}
\author{Daniel Cartin}
\email{cartin@naps.edu}
\affiliation{Naval Academy Preparatory School, 440 Meyerkord Avenue, Newport, Rhode Island 02841-1519}

\begin{abstract}

	The Standard Model of particle physics provides very accurate predictions of phenomena occurring at the sub-atomic level, but the reason for the choice of symmetry group and the large number of particles considered elementary, is still unknown. Along the lines of previous preon models positing a substructure to explain these aspects, Bilson-Thompson showed how the first family of elementary particles is realized as the crossings of braids made of three strands, with charges resulting from twists of those strands with certain conditions; in this topological model, there are only two distinct neutrino states. Modeling the particles as braids implies these braids must be the representation space of a Lie algebra, giving the symmetries of the Standard Model. In this paper, this representation is made explicit, obtaining the raising operators associated with the Lie algebra of $SU(5)$, one of the earliest grand unified theories. Because the braids form a group, the action of these operators are braids themselves, leading to their identification as gauge bosons. Possible choices for the other two families are also given. Although this realization of particles as braids is lacking a dynamical framework, it is very suggestive, especially when considered as a natural method of adding matter to loop quantum gravity.

\end{abstract}


\maketitle

\section{Introduction}
\label{intro}

Although it is a highly successful physical theory, aspects of the Standard model of particle physics remain mysterious, such as the choice of symmetry group $SU(3) \times SU(2) \times U(1)$, and the need for a relatively large number of fundamental particles. This has led to the notion that there must be a simpler underlying theory explaining these particular facts, continuing along the successful process of unification of physical theories begun by J. C. Maxwell in joining electricity and magnetism together. Many attempts, known as preon models, were developed to derive the properties of the Standard Model particles in terms of a smaller set of constituent particles. One such model is the Harari-Shupe model~\cite{rishon}, which posits two particles -- the $T$ particle with charge $+1/3$, and the $V$ particle with neutral charge -- and assumes all known particles are made of triplets of the $T$ and $V$ in various combinations. Negatively charged particles are due to the inclusion of the $\overline{T}$ anti-particle; there is also a neutral anti-particle $\overline{V}$ for the $V$ particle. In addition, it was assumed that in these groupings, particles and anti-particles cannot appear in the same triplet.

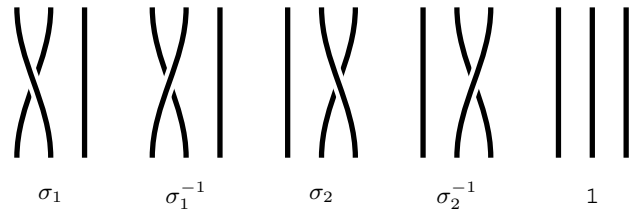
\begin{figure}[hbt]
\begin{tikzpicture}[
		strand/.style = {draw = white, ultra thick, double = black, double distance = 2 pt},
	]

	
	\def\a{0.45}		
	
	\begin{scope}[line width = 2 pt, rotate = 90]
	
	
		\draw (-1, 7*\a) cos (0, 6.5*\a) sin (1, 6*\a);
		\draw (-1, 5*\a) -- (1, 5*\a);
		\draw[strand] (-1, 6*\a) cos (0, 6.5*\a) sin (1, 7*\a);
		
	
		\draw (-1, 2*\a) cos (0, 2.5*\a) sin (1, 3*\a);
		\draw (-1, \a) -- (1, \a);
		\draw[strand] (-1, 3*\a) cos (0, 2.5*\a) sin (1, 2*\a);
		
	
		\draw (-1, -\a) -- (1, -\a);
		\draw (-1, -2*\a) cos (0, -2.5*\a) sin (1, -3*\a);
		\draw[strand] (-1, -3*\a) cos (0, -2.5*\a) sin (1, -2*\a);
		
		
		\draw (-1, -5*\a) -- (1, -5*\a);
		\draw (-1, -7*\a) cos (0, -6.5*\a) sin (1, -6*\a);
		\draw[strand] (-1, -6*\a) cos (0, -6.5*\a) sin (1, -7*\a);
		
	
		\draw (-1, -9*\a) -- (1, -9*\a);
		\draw (-1, -10*\a) -- (1, -10*\a);
		\draw (-1, -11*\a) -- (1, -11*\a);
	
	\end{scope}
	
	
	\node at (-6*\a, -1.5) {$\sigma_1$};
	\node at (-2*\a, -1.5) {$\sigma^{-1}_1$};
	\node at (2*\a, -1.5) {$\sigma_2$};
	\node at (6*\a, -1.5) {$\sigma^{-1}_2$};
	\node at (10*\a, -1.5) {$\mathbb{1}$};

\end{tikzpicture}
\caption{\label{crossings}Braids corresponding to each of the basic elements (and their inverses) in the braid group $B_3$, as well as the trivial (identity) braid $\mathbb{1}$.}
\end{figure}

Bilson-Thompson~\cite{Bil-Tho05} pointed out that, by using the Harari-Shupe model as an inspiration, the first family of Standard Model particles can be formulated in terms of braids, with charges represented by twists of the braid strands.  This uses braids with three strands, and two crossings of those strands. The basic generators of these braids are shown in Figure \ref{crossings}, along with the trivial braid; products of these generators, up to certain additional relations imposed on the products, form the braid group $B_3$~\cite{Mur-Kur99}. Charges are represented by twists on each strand: a $+e/3$ charge would be a full $2 \pi$ twist of the strand in one direction, while a $-e/3$ charge would be a twist in the other, and zero charge as a twist-free strand. The total charge of the particle in this realization would be the sum of the charges on all three strands. The Harari-Shupe criterion of only particles or anti-particles appearing in the same triplet now becomes the condition that any twists existing on the strands of a given braid are only in one direction, i.e. all positive or all negative. Particle interactions become the combination or decomposition of braids, using the rules of the braid group $B_3$. One advantage of the Bilson-Thompson model is that it provides a possible explanation of why there are only two neutrino states in a given family -- the left-handed neutrino and the right-handed anti-neutrino -- because acting with the parity operator gives braids identical to those neutrinos already known. Although there is not an exact correspondence between the two ideas, in the context of the rishon model, this can be thought of as making the $V$ and $\overline{V}$ particles identical in the braid realization.

In addition, this topological model of the particles fits naturally into the context of loop quantum gravity~\cite{Bil-Hac12}. This theory of quantum geometry uses spin networks -- graphs where the edges are labeled with group representations. When a cosmological constant is present, these edges must be represented by ribbons~\cite{Lambda}, allowing the possibility of the twists given in the Bilson-Thompson model. The dynamics in loop quantum gravity are generated by local moves on the graph, such as expansion and contraction of nodes, as well as exchanges of edges at a given node. A theory of quantum gravity such as this would provide both the context for why certain realizations of particles as braids are possible, and others are not, as well as the needed dynamical framework for the evolution and interaction of these realizations. In addition, the representations labelling the edges may have some relevance in why, as seen in this work, certain braids can be associated with gauge bosons of particular groups. It is certainly possible there is another formulation of the elementary particles equivalent to the topological preon model of Bilson-Thompson, more natural in the context of some overarching theory, but for the remainder of this paper, the choice of using three-strand braids with charges as twists is used.

This paper takes the ideas of Bilson-Thompson and adapts it to the minimal $SU(5)$ grand unified theory (GUT)~\cite{Geo-Gla74}, one of the earliest attempts at combining the color and electroweak groups into a single group. The placement of the particle realizations for the first family of Standard Model particles in the representations of $SU(5)$ given in this GUT are spelled out, along with the raising operators associated with the simple roots of the $SU(5)$ Lie algebra. Since the braids used are part of the braid group $B_3$, these raising operators (and their inverses, the lowering operators) will be braids themselves, allowing a straightforward realization of most of the gauge bosons in the same manner as the matter fermions. Although Bilson-Thompson did choose a representation of the Standard model gauge bosons as braids, it was not shown how these choices fit into the Lie algebra structure. In addition, the choice of using $SU(5)$ representations, rather than those of the Standard Model group $SU(3) \times SU(2) \times U(1)$, is used to see if these realizations of particles as braids has any explanatory power in why the elementary particles fit so easily into the representations of $SU(5)$. As mentioned above, these braid ideas already ``solve'' one of the problems seen in the $SO(10)$ GUT, that is having twice as many neutrino states, compared to what is currently observed. On the other hand, the minimal $SU(5)$ theory is experimentally ruled out, because of its prediction for proton decay; looking at the braid realizations of the gauge bosons responsible for these interactions may provide some insight into possible fixes for this issue.

The plan of the paper is the following. Before explicitly showing a possible set of particle realizations, the charge and parity operators $C$ and $P$, respectively, are defined in Section \ref{braid-ops}. These operations are used in Section \ref{particles} to give a set of particle realizations, as well as the operations corresponding to the raising operators of the Lie algebra for $SU(5)$. This section also shows why a choice of parity operator $P$ different than that chosen by Bilson-Thompson is needed to have a consistent set of operators and particle realizations. The necessary restriction on the properties of the parity operator shows the construction presented in this paper is not trivial. After a choice for the first particle family is exhibited, possible choices for the other two families are given, motivated by a restriction on the braids used for the particle realizations. The paper concludes in Section \ref{conclusions} with a discussion of this model of elementary particles as braids, in particular the aspects of the symmetry group, spin statistics and some comments on these particle realizations and the $SO(10)$ GUT.

\section{Braid operations}
\label{braid-ops}

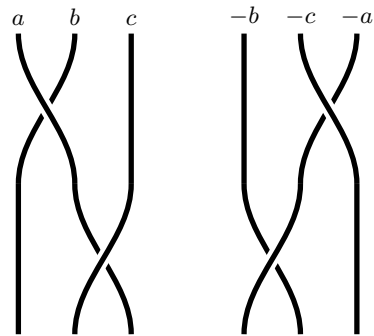
\begin{figure}
\begin{tikzpicture}[
		strand/.style = {draw = white, ultra thick, double = black, double distance = 2 pt},
	]

	
	\def\a{0.375}		
	
	\begin{scope}[line width = 2pt, rotate = 90]	
	
	
		\draw (-2, 6*\a) -- (0, 6*\a);				
		\draw (-2, 2*\a) cos (-1, 3*\a) sin (0, 4*\a);
		\draw[strand] (-2, 4*\a) cos (-1, 3*\a) sin (0, 2*\a);
		
		\draw (-2, -6*\a) -- (0, -6*\a);				
		\draw (-2, -4*\a) cos (-1, -3*\a) sin (0, -2*\a); 
		\draw[strand] (-2, -2*\a) cos (-1, -3*\a) sin (0, -4*\a);
		
	
		\draw (0, 6*\a) cos (1, 5*\a) sin (2, 4*\a);		
		\draw (0, 2*\a) -- (2, 2*\a);
		\draw[strand] (0, 4*\a) cos (1, 5*\a) sin (2, 6*\a);
		
		\draw (0, -2*\a) -- (2, -2*\a);				
		\draw (0, -4*\a) cos (1, -5*\a) sin (2, -6*\a);
		\draw[strand] (0, -6*\a) cos (1, -5*\a) sin (2, -4*\a);
	
	\end{scope}
	
	
	\node at (-6*\a, 2) [above] {$a$};				
	\node at (-4*\a, 2) [above] {$b$};
	\node at (-2*\a, 2) [above] {$c$};
	
	\node at (2*\a, 2) [above] {$-b$};				
	\node at (4*\a, 2) [above] {$-c$};
	\node at (6*\a, 2) [above] {$-a$};

\end{tikzpicture}
\caption{\label{structure}The braid and charge structures corresponding to the state $\sigma_1 \sigma^{-1} _2 [a, b, c]$ and its inverse $\sigma_2 \sigma^{-1} _1 [-b, -c, -a]$. The permutation of the charge structure in the inverse is needed because of how the strands change position, moving from top to bottom in the braids $\sigma_1 \sigma_2$. Note that the product $\sigma_1 \sigma_2$, for example, is represented as placing the $\sigma_1$ crossing on top of the $\sigma_2$ crossing. The three variables $a, b, c \in \{-1, 0, 1 \}$ represent the charge structure of the particle realizations.}
\end{figure}

The original model by Bilson-Thompson has two features associated with each particle in the Standard Model -- the braid used for the particle, and the positive and negative charges placed on each strand of the braid. The former is referred to here as the {\it braid structure} of the particle, and the latter as the {\it charge structure}. For a given particle realization where the braid structure is, e.g. $\sigma_1 \sigma^{-1} _2$ and the charge structure has a $+2\pi$ twist equivalent to $+e/3$ on the first strand, no twist on the middle strand, and a $-2\pi$ twist equivalent to $-e/3$ on the third strand, the notation used for this particle state is $\sigma_1 \sigma^{-1} _2 [+1, 0, -1]$; however, this example is only for illustrative purposes, since it violates the Bilson-Thompson rule against charges of opposite sign in the same braid mentioned in Section \ref{intro}. To avoid excessive notation, this paper uses the symbols $\sigma_1$ and $\sigma_2$ to denote both the members of the abstract braid group $B_3$, as well as the braid structure for the physical particle states of the Standard Model; the difference between the two should be obvious from the context. Note that this notation assumes the charge is placed at the top of each strand, as shown in Figure \ref{structure}.

The multiplication $\sigma_a \sigma_b$ of two braids $\sigma_a$ and $\sigma_b$ is realized as first completing the braid crossing for $\sigma_a$, then that of $\sigma_b$. In other words, the first braid is placed ``on top'' of the second. The left-hand picture in Figure \ref{structure} shows the multiplication $\sigma_1 \sigma^{-1} _2$, where the crossing $\sigma_1$ of the left two strands is done above the crossing $\sigma^{-1} _2$ of the right two strands. In addition, Figure \ref{structure} shows how the braid $\sigma_1 \sigma^{-1} _2$ acts like a permutation of the charge labels -- moving from the top of the braid to the bottom, the charges are rearranged
\begin{equation}
\label{s1s2}
[a, b, c] \to [b, c, a]
\end{equation}
Note that this permutation does not change as long as the crossing of the left two strands is at the top, and that of the right two strands is at the bottom, so that $\sigma^{-1}_1 \sigma_2$ would give the same result. On the other hand, the braid $\sigma_2 \sigma^{-1} _1$ (and similar crossing patterns) gives a permutation of
\begin{equation}
\label{s2s1}
[a, b, c] \to [c, a, b]
\end{equation}
These permutations show why the realizations $\sigma_1 \sigma^{-1} _2 [a, b, c]$ and $\sigma_2 \sigma^{-1} _1 [-b, -c, -a]$, shown in Figure \ref{structure}, are inverses. Placing either of these two braids on top of the other -- i.e. multiplying the two in either order -- it is obvious that the strands can be moved around until the combined braid is the trivial braid $\mathbb{1}$. In addition, the placement of the three charges on $\sigma_2 \sigma^{-1} _1 [-b, -c, -a]$ is exactly what is needed to cancel the charge structure in $\sigma_1 \sigma^{-1} _2 [a, b, c]$, either by moving the charges $[-b, -c, -a]$ upward through $\sigma_1 \sigma^{-1} _2$ via the reverse of the permutation (\ref{s1s2}) or downward through $\sigma^{-1} _2 \sigma_1$ using the permutation (\ref{s2s1}) to meet the top of $\sigma_1 \sigma^{-1} _2$.

The $C$ operation maps a particle to its anti-particle partner; in terms of braids, it results in the inverse of the original particle, both in its braid and charge structures. This mapping acts like a top-down reflection of the braid, thus reversing the direction of the twists on each strand, and placing them at the bottom of the braid. The inverse of the braid structure gives
\begin{equation}
\label{C-def}
	C(\sigma_1) = \sigma^{-1} _1 \qquad
	C(\sigma_2) = \sigma^{-1} _2
\end{equation}
and
\begin{equation}
\label{C-order}
	C(\sigma_a \sigma_b) = C(\sigma_b) C(\sigma_a)
\end{equation}
The charge structure of $C(\sigma_a \sigma_b)$ must take into account the permutation of the strands as discussed above, in order to move them back to the top of the braid strands after the top-down reflection. As examples, for the braids $\sigma_1 \sigma^{-1} _2$ and $\sigma_2 \sigma^{-1} _1$, acting with $C$ gives
\begin{subequations}
\label{C-permute}
\begin{eqnarray}
	C(\sigma_1 \sigma^{-1} _2 [a, b, c]) &=& \sigma_2 \sigma^{-1} _1 [-b, -c, -a]	\\
	C(\sigma_2 \sigma^{-1} _1 [a, b, c]) &=& \sigma_1 \sigma^{-1} _2 [-c, -a, -b]
\end{eqnarray}
\end{subequations}
where the charge structure of the braid after being acted on by $C$ is what is necessary to cancel out the permutations (\ref{s1s2}) and (\ref{s2s1}). When a braid and its inverse are multiplied, the resulting braid is the identity braid $\mathbb{1}$, represented as three strands with no crossings.

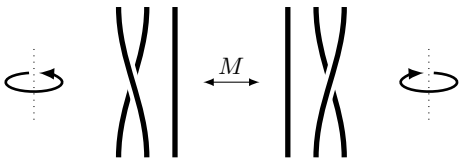
\begin{figure}[hbt]
\begin{tikzpicture}[
		strand/.style = {draw = white, ultra thick, double = black, double distance = 2 pt},
		> = latex,
	]

	
	\def\a{0.375}		
	
	\begin{scope}[line width = 2 pt, rotate = 90]
	
	
		\draw (-1, 2*\a) -- (1, 2*\a);
		\draw (-1, 4*\a) cos (0, 3.5*\a) sin (1, 3*\a);
		\draw [strand] (-1, 3*\a) cos (0, 3.5*\a) sin (1, 4*\a);
	
	
		\draw (-1, -2*\a) -- (1, -2*\a);
		\draw (-1, -4*\a) cos (0, -3.5*\a) sin (1, -3*\a);
		\draw [strand] (-1, -3*\a) cos (0, -3.5*\a) sin (1, -4*\a);
	
	\end{scope}
		
	
		\draw [dotted] (-7*\a, -0.5) -- (-7*\a, 0.5);
		\draw [->, line width = 1 pt] (-7*\a, 0) ++ ({-\a * sin(10)}, {0.125 * cos(10)}) arc (100 : 440 : {\a} and 0.125);
		
	
		\draw [dotted] (7*\a, -0.5) -- (7*\a, 0.5);
		\draw [->, line width = 1 pt] (7*\a, 0) ++ ({\a * sin(10)}, {0.125 * cos(10)}) arc (80 : -260 : {\a} and 0.125);

	
	\draw[<->] (-\a, 0) -- node [midway, above] {$M$} (\a, 0);

\end{tikzpicture}
\caption{\label{mirror}The mirror operation $M$ acting on a braid, resulting in a left-right reflection. Here is shown $M(\sigma_1) = \sigma^{-1} _2$. If the charge structure is represented by twists in the strands (as in the original Bilson-Thompson model), then $M$ changes the charge structure to its negative, as shown by the circular arrows, as well as in reverse order, i.e. $M(\sigma_1 [a, b, c]) = \sigma^{-1} _2 [-c, -b, -a]$; the three variables $a, b, c \in \{-1, 0, 1 \}$ represent the charge structure of the particle realizations.}
\end{figure}

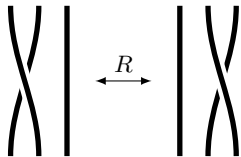
\begin{figure}
\begin{tikzpicture}[
		strand/.style = {draw = white, ultra thick, double = black, double distance = 2 pt},
		> = latex,
	]

	
	\def\a{0.375}		
	
	\begin{scope}[line width = 2 pt, rotate = 90]
	
	
		\draw (-1, 2*\a) -- (1, 2*\a);
		\draw(-1, 4*\a) cos (0, 3.5*\a) sin (1, 3*\a);
		\draw [strand] (-1, 3*\a) cos (0, 3.5*\a) sin (1, 4*\a);
	
	
		\draw (-1, -2*\a) -- (1, -2*\a);
		\draw (-1, -3*\a) cos (0, -3.5*\a) sin (1, -4*\a);
		\draw [strand] (-1, -4*\a) cos (0, -3.5*\a) sin (1, -3*\a);
	
	\end{scope}

	
	\draw[<->] (-\a, 0) -- node [midway, above] {$R$} (\a, 0);

\end{tikzpicture}
\caption{\label{reversal}The reversal operation $R$ acting on a braid, resulting in a braid seen from ``the other side'' of the page. Here, $R(\sigma_1) = \sigma_2$. The signs of the charge structure -- the twists on each strand -- are unaffected by $R$, although their order is reversed, so $R(\sigma_1 [a, b, c]) = \sigma_2 [c, b, a]$; the three variables $a, b, c \in \{-1, 0, 1 \}$ represent the charge structure of the particle realizations.}
\end{figure}

In order to define a parity operator $P$, it is necessary to consider two additional operations first. The mirror operation $M$ is defined as mapping a braid to its left-right mirror image, as shown in Figure \ref{mirror}. For the braid structure, this operation gives
\begin{equation}
\label{M-def}
	M(\sigma_1) = \sigma^{-1} _2 \qquad
	M(\sigma_2) = \sigma^{-1} _1
\end{equation}
and
\begin{equation}
\label{M-order}
	M(\sigma_a \sigma_b) = M(\sigma_a) M(\sigma_b)
\end{equation}
If the charge structure is kept with the same sign, but the charges are mirrored along with the strands they reside on, this is the parity operation defined by Bilson-Thompson. However, for the definition here, the mirror operation also affects the charge structure. Since in the original model, the charges were represented as twists on each strand, taking the left-right mirror will reverse the rotation of these twists as well; in Figure \ref{mirror}, this is represented by a flip in the direction of the twists on either side of the braids. Thus, for this paper, the mirror operation $M$ is defined as acting on the charge structure as
\begin{equation}
\label{M-permute}
	M([a, b, c]) = [-c, -b, -a]
\end{equation}
The last operation needed to define the parity operator $P$ is the reversal operation $R$, shown in Figure \ref{reversal}. The intuition for this operation is to look at the ``back side'' of the braid, i.e. from the other side of the page. For example, if looking at the left-hand braid $\sigma_1$ in Figure \ref{reversal}, from the other side, the middle strand passes over the right-hand strand, rather than under the left-hand strand, leading to the crossing $\sigma_1$ on one side appearing as a $\sigma_2$ crossing on the opposite side of the braid. The operator $R$ acting on the braid structure gives
\begin{equation}
\label{R-def}
	R(\sigma_1) = \sigma_2 \qquad
	R(\sigma_2) = \sigma_1
\end{equation}
and
\begin{equation}
\label{R-order}
	R(\sigma_a \sigma_b) = R(\sigma_a) R(\sigma_b)
\end{equation}
Twists on each strand retain the same parity, so the signs of the charge structure are unaffected; however, their order is reversed, along with the order of the strands. This gives
\begin{equation}
\label{R-permute}
	R([a, b, c]) = [c, b, a]
\end{equation}
Because $R$ does not change the physical realization of the particle as a braid, it is assumed that all particle identifications are invariant under this reversal operation. As will be seen below, this is the assumption reducing the neutrino states down to two independent realizations.  Note all of these operations square to the identity, with $C^2 = M^2 = R^2 = \mathbb{1}$, and they also commute, with
\[
	[C, M] = [C, R] = [M, R] = 0
\]

The mirror operation $M$ changes the charge of the particle, so this by itself is insufficient for a parity operator, since the left- and right-handed versions of a given particle should have the same charge. Thus, a charge inversion operation must be included also, so we define the parity operator in terms of the charge inversion $C$ and mirror operation $M$ as $P = RCM$. This choice is perhaps most naturally justified by looking at the action of $CP = RM$ on a particle; by using the definitions of $R$ and $M$ given previously, as an example, $CP(\sigma_1 \sigma^{-1} _2 [a, b, c]) = \sigma^{-1}_1 \sigma_2 [-a, -b, -c]$. As one would expect for $CP$, the charge structure of the resulting particle has the opposite sign as the original, while each crossing in the braid structure is mapped to its inverse, although the order of the crossings remains the same. This definition of $P$ also allows all particles to be consistently placed in the representations of $SU(5)$, as shown in Section \ref{particles}. Because of the combination of operators inside $P$, the crossings themselves are individually unaffected, so
\begin{equation}
\label{P-def}
	P(\sigma_1) = \sigma_1 \qquad
	P(\sigma_2) = \sigma_2
\end{equation}
but the presence of the $C$ operation inside $P$ gives a change in ordering:
\begin{equation}
\label{P-order}
	P(\sigma_a \sigma_b) = P(\sigma_b) P(\sigma_a)
\end{equation}
In addition, the $C$ operation inside $P$ means the effect of the parity operator on the charge structure is dependent on the braid structure. In particular, for the choices $\sigma_1 \sigma^{-1} _2$ and $\sigma_2 \sigma^{-1} _1$, acting with $P$ gives
\begin{subequations}
\label{P-permute}
\begin{eqnarray}
	P(\sigma_1 \sigma^{-1} _2 [a, b, c]) &=& \sigma^{-1} _2 \sigma_1 [b, c, a]	\\
	P(\sigma_2 \sigma^{-1} _1 [a, b, c]) &=& \sigma^{-1} _1 \sigma_2 [c, a, b]
\end{eqnarray}
\end{subequations}
As an example of these definitions, the results of acting on a particle state $\sigma_1 \sigma^{-1} _2 [a, b, c]$ with the charge and parity operators is shown in Figure \ref{CP}.

\begin{figure}[hbt]
\begin{tikzpicture}[> = latex]

	
	\node (p-L) {$\sigma_1 \sigma^{-1} _2 [a, b, c]$};
	\node (p-R) [right = 6 em of p-L] {$\sigma^{-1} _2 \sigma_1 [b, c, a]$};
	\node (pbar-L) [below = 2 em of p-R] {$\sigma^{-1} _1 \sigma _2 [-a, -b, -c]$};
	\node (pbar-R) [below = 2 em of p-L] {$\sigma_2 \sigma^{-1} _1 [-b, -c, -a]$};
	
	
	\begin{scope}[<->]
	
		\draw (p-L) -- node [midway, above] {$P$} (p-R);
		\draw (p-L) -- node [midway, left] {$C$} (pbar-R);
		\draw (pbar-L) -- node [midway, right] {$C$} (p-R);
		\draw (pbar-R) -- node [midway, above] {$P$} (pbar-L);
	
	\end{scope}

\end{tikzpicture}
\caption{\label{CP}Charge inversion $C$ and parity inversion $P$ operators acting on the braids $\sigma_1 \sigma^{-1} _2$ and $\sigma^{-1} _1 \sigma_2$; the three variables $a, b, c \in \{-1, 0, 1 \}$ represent the charge structure of the particle realizations.}
\end{figure}

\section{Realizations of particles as braids}
\label{particles}

Now that charge and parity operators have been defined on the particle realizations, the particles of the Standard Model can be determined. However, in this work the Lie algebra structure of these particles is also imposed on these choices, leading to a large restriction in the possible realizations. These structures are now laid out. Recall that SU(5) is a rank 4 Lie algebra, with four simple roots $\alpha^i$. In the spontaneous symmetry breaking process $SU(5) \to SU(3) \times SU(2) \times U(1)$, $\alpha^1$ and $\alpha^2$ become associated with the raising operators of the color $SU(3)$ group, while $\alpha^4$ is associated with the electroweak $U(1)$ and $\alpha^3$ is related to non-conservation of baryon and lepton number. The $SU(5)$ GUT places all the first family of particles into four representations; Figure \ref{10-rep} shows the \rep{10} representation, containing left-handed particles, along with the actions of the raising operators associated with the four simple roots $\alpha^i$. From now on, $\alpha^i$ is used to denote both the simple root and the raising operator for that root.

\begin{figure}[hbt]
\begin{tikzpicture}[> = latex]


	\node (ubar_3) {$\overline{u}^3 _L$};
	\node (ubar_2) [below = 2 em of ubar_3]{$\overline{u}^2 _L$};
	\node (ubar_1) [below left = 2 em of ubar_2] {$\overline{u}^1 _L$};
	\node (u_1) [below right = 2 em of ubar_2] {$u^1 _L$};
	\node (u_2) [below = 2 em of ubar_1] {$u^2 _L$};
	\node (u_3) [below = 2 em of u_2] {$u^3 _L$};
	\node (d_1) [below = 2 em of u_1] {$d^1 _L$};
	\node (d_2) [below = 2 em of d_1] {$d^2 _L$};
	\node (d_3) [below right = 2 em of u_3] {$d^3 _L$};
	\node (epos) [below = 2 em of d_3] {$e^+ _L$};
	
	
	\begin{scope}[->]
	
		\draw (ubar_2) -- node [midway, right] {$\alpha_2$} (ubar_3);
		\draw (ubar_1) -- node [midway, above left] {$\alpha^1$} (ubar_2);
		\draw (u_1) -- node [midway, above right] {$\alpha^3$} (ubar_2);
		\draw (u_2) -- node [midway, left] {$\alpha^3$} (ubar_1);
		\draw (u_2) -- node [midway, fill = white] {$\alpha^1$} (u_1);
		\draw (d_1) -- node [midway, right] {$\alpha^4$} (u_1);
		\draw (u_3) -- node [midway, left] {$\alpha^2$} (u_2);
		\draw (d_2) -- node [midway, right] {$\alpha^1$} (d_1);
		\draw (d_2) -- node [midway, fill = white] {$\alpha^4$} (u_2);
		\draw (d_3) -- node [midway, below left] {$\alpha^4$} (u_3);
		\draw (d_3) -- node [midway, below right] {$\alpha^2$} (d_2);
		\draw (epos) -- node [midway, right] {$\alpha^3$} (d_3);
	
	\end{scope}
	
	

\end{tikzpicture}
\caption{\label{10-rep}The particle content of the \rep{10} representation in the SU(5) GUT, along with the action of the raising operators associated with the simple roots $\alpha^i$ of SU(5). The upper indices for each of the quark states is its color index in SU(3); all particles in this representation are left-handed, denoted by the $L$ subscript.}
\end{figure}

One assumption used in the realization given here is that all quarks of a given type and parity (e.g. left-hand up quarks) have the same braid structure. This puts a restriction on the allowable form of the raising operations associated with the simple roots $\alpha^i$. Since $\alpha^1$ and $\alpha^2$ change only the color index of a given type of quark, the action of these operators cannot change the braid structure, only the charge structure. This means that these two operators have the trivial braid structure $\mathbb{1}$, and the charge structure raises or lowers the charge on a given strand by one unit, in order to change the position of the charges on the strands for the quark states. However, starting with $u^2 _L$ as an example, $\alpha^3$ must change both the braid and charge structure, since acting on $u^2 _L$ with $\alpha^3$ leads to $\overline{u}^1 _L$, which cannot have the same structure as $u^1 _L$ or $u^2 _L$ (the numerical superscript for all quark states is the color $SU(3)$ index; the $L$ subscript indicates the particles are left-handed). If this were not the case, then each $\overline{u}^i _L$ would be the anti-particle for one of the $u^i _L$, as well as its right-handed partner $u^i _R$ under charge conjugation. In addition, $\alpha^3$ and $\alpha^4$ must be inverses of each other, at least for their braid structure, as seen by the following. With $\alpha^1$ and $\alpha^2$ acting trivially as braids, then any possible chain of raising operators taking the $e^+ _L$ particle to the $\overline{u}^2 _L$ particle in Figure \ref{10-rep} has $\alpha^3$ acting twice and $\alpha^4$ acting once; if $\alpha^3$ and $\alpha^4$ were not braid inverses, this process leads to braids with more than two strand crossings.

For the \rep{10} representation of the $SU(5)$ GUT, the choice is made to act on the particle realizations to the right with the raising operators. In this representation, both the left-handed up quarks $u^i _L$ and the left-handed anti-up quarks $\overline{u}^i _L$ appear; making sure that both are in the representation, and are still related by the action of the $CP$ operator, restricts the choices of realizations. Specifically, note that the $SU(3)$ action on these quarks results in $u^2 _L = u^3 _L \cdot \alpha^2$ and $u^1 _L = u^2_L \cdot \alpha^1$. In addition, the same raising operators $\alpha^1$ and $\alpha^2$ act on the anti-up quarks to give $\overline{u}^2 _L = \overline{u}^1 _L \cdot \alpha^1$ and $\overline{u}^3 _L = \overline{u}^2 _L \cdot \alpha^2$. Combining these together, with the fact that $\overline{u}^i _L = CP(u^i _L)$ gives
\begin{subequations}
\label{ubar-eqns-01}
\begin{eqnarray}
	\overline{u}^2 _L &=& CP(u^2 _L \cdot \alpha^1) \cdot \alpha^1		\\
	\overline{u}^3 _L &=& CP(u^3_L \cdot \alpha^2) \cdot \alpha^2
\end{eqnarray}
\end{subequations}
As given by Bilson-Thompson, the parity operator is a left-right mirror symmetry of the braid structure, without changing the sign of the charge structure (just keeping each of the three on the same reflected strand). With this choice, the combined operator $CP_{BT}$ has an action $CP_{BT} (u^i _L \cdot \alpha^j) = CP_{BT} (\alpha^j) \cdot CP_{BT} (u^i _L)$, with the reversal of the braid order coming from the charge inversion operator $C$. Therefore, the above equalities (\ref{ubar-eqns-01}) in the Bilson-Thompson model are now
\begin{subequations}
\label{ubar-BT-eqns-01}
\begin{eqnarray}
	\overline{u}^2 _L &=& CP_{BT} (\alpha^1) \cdot \overline{u}^2 _L \cdot \alpha^1	\\
	\overline{u}^3 _L &=& CP_{BT} (\alpha^2) \cdot \overline{u}^3 _L \cdot \alpha^2
\end{eqnarray}
\end{subequations}

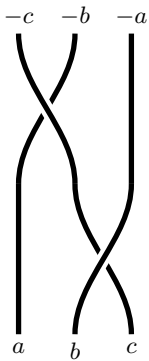
\begin{figure}[hbt]
\begin{tikzpicture}[
		strand/.style = {draw = white, ultra thick, double = black, double distance = 2 pt},
	]

	
	\def\a{0.375}		
	
	\begin{scope}[line width = 2pt, rotate = 90]	
	
	
		\draw (-2, 6*\a) -- (0, 6*\a);	
		\draw (-2, 2*\a) cos (-1, 3*\a) sin (0, 4*\a);
		\draw[strand] (-2, 4*\a) cos (-1, 3*\a) sin (0, 2*\a);
		
	
		\draw (0, 6*\a) cos (1, 5*\a) sin (2, 4*\a);
		\draw (0, 2*\a) -- (2, 2*\a);
		\draw[strand] (0, 4*\a) cos (1, 5*\a) sin (2, 6*\a);
	
	\end{scope}
	
	
	\node at (-6*\a, 2) [above] {$-c$};
	\node at (-4*\a, 2) [above] {$-b$};
	\node at (-2*\a, 2) [above] {$-a$};
	
	\node at (-6*\a, -2) [below] {$a$};
	\node at (-4*\a, -2) [below] {$b$};
	\node at (-2*\a, -2) [below] {$c$};
	
\end{tikzpicture}
\caption{\label{BT-cancel}Example of cancellation needed to make the Bilson-Thompson operator $P_{BT}$ work consistently with the Lie algebra structure of the Standard Model. This illustration represents graphically the operations in (\ref{ubar-BT-eqns-01}) using the choice $\sigma^{-1} _1 \sigma_2$ for the charge structure of $\overline{u}^i _L$. Here the charges $\{a, b, c\}$ at the bottom come from the action of $\alpha^i$, while those of $\{-a, -b, -c\}$ at the top come from the action of $CP_{BT} (\alpha^i)$.}
\end{figure}

As argued previously, the braid structures of $\alpha^1$ and $\alpha^2$ are both the trivial identity braid, so in order to have these braid consistency relations be satisfied, the charge structures for $CP_{BT} (\alpha^i)$, placed at the top of the braid, and $\alpha^i$, placed at the bottom, must cancel out. This is impossible to do unless two of the three charges on the $\alpha^i$ strands are the same. To see this, recall that $C$ acting on the trivial braid $\mathbb{1}$ will not change the braid structure, but it will map the charge structure to its inverse. Thus,
\[
	C(\mathbb{1} [a, b, c]) = \mathbb{1} [-a, -b, -c]
\]
The parity operator $P_{BT}$ chosen by Bilson-Thompson does not alter the sign of the charges, but does reverse their order, so
\[
	CP_{BT} (\mathbb{1} [a, b, c]) = \mathbb{1} [-c, -b, -a]
\]
The requirement that the relations (\ref{ubar-BT-eqns-01}) means that these charges $a, b, c$ must cancel out, but after they are permuted by the crossings in the braids $\overline{u}^2 _L$ and $\overline{u}^3 _L$. Specifically, suppose we choose the braid structure of $\overline{u}^i _L$ to be $\sigma^{-1} _1 \sigma_2$. Then, as seen in Figure \ref{BT-cancel}, this means that $a = b$, while $c$ is undetermined.

However, this requirement means it is impossible to have all three up quark states have charge structures as permutations of $\{+1, +1, 0 \}$. Continuing with the example, if we let $\overline{u}^1 _L = \sigma^{-1} _1 \sigma_2 [a', b', c']$, then by acting with $\alpha^i = \mathbb{1} [a^i, a^i, c^i], i=1, 2$, we have
\begin{eqnarray*}
	\overline{u}^2 _L &=& \overline{u}^1 _L \cdot \alpha^1 = \sigma^{-1} _1 \sigma_2 [a' + c^1, b' + a^1, c' + a^1]	\\
	\overline{u}^3 _L &=& \overline{u}^2 _L \cdot \alpha^2 \\
		&=& \sigma^{-1} _1 \sigma_2 [a' + c^1 + c^2, b' + a^1 + a^2, c' + a^1 + a^2]
\end{eqnarray*}
where the permuted actions of $a^i$ and $c^i$ come from shifting the charges of $\alpha^i$ from the bottom to the top of the braid $\sigma^{-1} _1 \sigma_2$ for $\overline{u}^j _L$ (this is the reverse of the charge structure permutation shown in (\ref{s1s2})). For any choice of $\{a^i, b^i, c^i\}$ as a permutation of $\{+1, +1, 0\}$ for the particle $\overline{u}^1 _L$, there are no consistent values for the four parameters $a^1, c^1, a^3, c^3$ giving the charge structures for the other two particles $\overline{u}^2 _L$ and $\overline{u}^3 _L$ as the other two permutations of $\{+1, +1, 0\}$. This issue does not go away if the action of the $\alpha^i$ is to the left rather than the right -- there is still the problem of $\alpha^i$ and $CP(\alpha^i)$ acting on opposite sides of the particle realization.
 
On the other hand, if as defined in this paper, the $CP$ operator acting on a product of braids preserves the order of this product because of the relations (\ref{C-order}) and (\ref{P-order}) on $C$ and $P$, respectively, then solutions to this issue exist. In particular, with
\[
	\overline{u}^2 _L = CP(u^2 _L \cdot \alpha^1) \cdot \alpha^1 = \overline{u}^2 _L \cdot CP(\alpha^1) \cdot \alpha^1
\]
and similarly for $\overline{u}^3 _L$, then the particles are consistent if $CP(\alpha^1) = (\alpha^1)^{-1}$ and $CP(\alpha^2) = (\alpha^2)^{-1}$, i.e.
\begin{equation}
\label{alpha-parity}
	P(\alpha^i) = \alpha^i \qquad i=1, 2
\end{equation}
Using the definition of $P = RCM$ given previously,
\[
	P(\mathbb{1} [a, b, c]) = \mathbb{1}[a, b, c]
\]
Thus, for any choice of $\alpha^1$ and $\alpha^2$, the particle realizations in the \rep{10} representation will be consistent. A particular choice of raising operators associated with each simple root of $SU(5)$ is given in Table \ref{alpha-def}.

\begin{table}[hbt]
\begin{tabular}{cc}
Simple root	& Raising operator							\\
\hline
$\alpha^1$	& $\mathbb{1} [+1, 0, -1]$						\\
$\alpha^2$	& $\mathbb{1} [-1, +1, 0]$						\\
$\alpha^3$	& $\sigma_2 \sigma^{-2} _1 \sigma_2 [-1, -2, -1]$		\\
$\alpha^4$	& $\sigma^{-1} _2 \sigma^2 _1 \sigma^{-1} _2 [+1, +1, +1]$	\\
\end{tabular}
\caption{\label{alpha-def}Raising operators associated with each of the simple roots of SU(5) for the first generation of Standard Model particles; the corresponding lowering operators are the inverses of those given in the table. The braid structure $\mathbb{1}$ is the trivial (identity) braid. Note that all four raising operators are invariant under the parity operator $P$. The identification of these operators as particles is given in Table \ref{adjoint}.}
\end{table}

One nice consequence of these definitions of the raising operators is in how the $\alpha^i$ act on the braids in each representation. Remember that for the \rep{10} representation, the raising operators acted on the right, e.g. $p_L \to p' _L = p_L \cdot \alpha^i$. Since the particles in the \rep{5} representation are obtained from those in \rep{10} by acting with the parity operator $P$, then
\[
	p' _R  = P(p' _L) = P(p_L \cdot \alpha^i) = P(\alpha^i) \cdot p_R
\]
Since the requirement for consistency in the \rep{10} representation led to the parity invariance (\ref{alpha-parity}) of $\alpha^1$ and $\alpha^2$, then the same raising operators used for \rep{10} also work for \rep{5}, as long as they act on the left instead of the right. By explicit computation, the operators $\alpha^3$ and $\alpha^4$ share this parity invariance, so all four operators act equally well on \rep{5}. The same logic works as well for the conjugate representations $\repbar{5}$ and $\repbar{10}$. For instance, in the $\repbar{5}$ representation, the raising operator acts as $\alpha^i : \overline{p}' _L \to \overline{p} _L$, i.e. in the reverse order to how the particles appear in the \rep{10}. Since the particles in this representation are obtained from the \rep{10} by acting with $CP$, then
\[
	\overline{p}' _L = CP(p'_L) = CP(p_L \cdot \alpha^i)
\]
As seen before, the action of $CP$ keeps the same order of the braid product, so that
\[
	\overline{p}' _L = \overline{p}_L \cdot CP(\alpha^i)
\]
Parity invariance of $\alpha^i$ gives that $CP(\alpha^i) = (\alpha^i)^{-1}$; thus by acting on the right with $\alpha^i$, one obtains
\[
	\overline{p}_L = \overline{p}'_L \cdot \alpha^i	
\]
Going through a similar logic, the particles in $\repbar{10}$ are acted on the left by the $\alpha^i$; this matches up with the idea that one representation and its conjugate have the Lie algebra operators act on the ``opposite side'' of each other.

To summarize the above discussion, a consistent realization of all the particles in the Standard Model is obtained by the following choices:

\begin{enumerate}

	\item Choose a braid structure for the left-handed positron $e^+ _L$; the charge structure must be $[+1, +1, +1]$.
	
	\item In terms of braid structure, $e^+ _L$ and $u^i _L$ must be the same, since $\alpha^3$ and $\alpha^4$ are inverses, at least for the braid structure. This means the braid structure of $u^i _L$ is fixed by the choice of $e^+ _L$; assigning distinct permutations of the charge structure $\{+1, +1, 0 \}$ to each of the three left-handed up quark particles fixes the operators $\alpha^1$ and $\alpha^2$.
	
	\item Since $\overline{u}^i _L$ is determined by $CP$ acting on $u^i _L$, the raising operator $\alpha^3$ is now obtained. From this, the realizations for all $d^i _L$ are also fixed (since $\alpha^1$ and $\alpha^2$ are already known), and so is the action of the operator $\alpha^4$.
	
	\item All the other three representations \rep{5}, $\repbar{5}$ and $\repbar{10}$ are determined by these choices, by using $P$, $CP$, and $C$, respectively, on the \rep{10} representation. The raising operators act on the right for the left-handed particle representations $\repbar{5}$ and \rep{10}, and on the left for the right-handed particle representations \rep{5} and $\repbar{10}$.

\end{enumerate}
The upshot of this is that there are only three inputs to determine all of the braid and charge structures -- the braid structure for $e^+ _L$ and the charge structures for two of the three left-handed up quarks, with the third being the remaining one not used previously. A complete set of particle realizations in shown in Table \ref{SM-part}. Note that, because the braids associated with the neutrino have zero charge, it is automatically true that acting on one of them with the parity operator, e.g. $P(\nu_L)$, automatically gives the same result as acting on its anti-matter partner with the reversal symmetry $R$, so $R(\overline{\nu}_R) = P(\nu_L)$. Because of the assumption that the physical content of the model is invariant under $R$, this gives only two independent neutrino states.

\begin{table}[hbt]
\begin{tabular}{cccc}
Particle			& Structure						& Particle			& Structure						\\
\hline
$u^1 _L$			& $\sigma_1 \sigma^{-1} _2 [0, +1, +1]$	& $u^1 _R$			& $\sigma^{-1} _2 \sigma_1 [+1, +1, 0]$	\\
$u^2 _L$			& $\sigma_1 \sigma^{-1} _2 [+1, 0, +1]$	& $u^2 _R$			& $\sigma^{-1} _2 \sigma_1 [0, +1, +1]$	\\
$u^3 _L$			& $\sigma_1 \sigma^{-1} _2 [+1, +1, 0]$	& $u^3 _R$			& $\sigma^{-1} _2 \sigma_1 [+1, 0, +1]$	\\
\hline
$\overline{u}^1 _L$	& $\sigma^{-1} _1 \sigma_2 [0, -1, -1]$	& $\overline{u}^1 _R$	& $\sigma_2 \sigma^{-1} _1 [-1, -1, 0]$	\\
$\overline{u}^2 _L$	& $\sigma^{-1} _1 \sigma_2 [-1, 0, -1]$	& $\overline{u}^2 _R$	& $\sigma_2 \sigma^{-1} _1 [0, -1, -1]$	\\
$\overline{u}^3 _L$	& $\sigma^{-1} _1 \sigma_2 [-1, -1, 0]$	& $\overline{u}^3 _R$	& $\sigma_2 \sigma^{-1} _1 [-1, 0, -1]$	\\
\hline
$d^1 _L$			& $\sigma_1^{-1} \sigma_2 [-1, 0, 0]$		& $d^1 _R$			& $\sigma_2 \sigma^{-1} _1 [0, 0, -1]$		\\
$d^2 _L$			& $\sigma_1^{-1} \sigma_2 [0, -1, 0]$		& $d^2 _R$			& $\sigma_2 \sigma^{-1} _1 [-1, 0, 0]$		\\
$d^3 _L$			& $\sigma_1^{-1} \sigma_2 [0, 0, -1]$		& $d^3 _R$			& $\sigma_2 \sigma^{-1} _1 [0, -1, 0]$		\\
\hline
$\overline{d}^1 _L$	& $\sigma_1 \sigma^{-1} _2 [+1, 0, 0]$	& $\overline{d}^1 _R$	& $\sigma^{-1} _2 \sigma_1 [0, 0, +1]$	\\
$\overline{d}^2 _L$	& $\sigma_1 \sigma^{-1} _2 [0, +1, 0]$	& $\overline{d}^2 _R$	& $\sigma^{-1} _2 \sigma_1 [+1, 0, 0]$	\\
$\overline{d}^3 _L$	& $\sigma_1 \sigma^{-1} _2 [0, 0, +1]$	& $\overline{d}^3 _R$	& $\sigma^{-1} _2 \sigma_1 [0, +1, 0]$	\\
\hline
$e^+ _L$			& $\sigma_1 \sigma^{-1} _2 [+1, +1, +1]$	& $e^+ _R$			& $\sigma^{-1} _2 \sigma_1 [+1, +1, +1]$	\\
$e^- _L$			& $\sigma^{-1} _1 \sigma_2 [-1, -1, -1]$	& $e^- _R$			& $\sigma_2 \sigma^{-1} _1 [-1, -1, -1]$	\\
$\nu_L$			& $\sigma_1 \sigma^{-1} _2 [0, 0, 0]$		& $\overline{\nu}_R$	& $\sigma_2 \sigma^{-1} _ 1 [0, 0, 0]$		\\
\hline
\end{tabular}
\caption{\label{SM-part}Braid and charge structures for all particles in the first family of the Standard Model; the numerical superscripts for the quark states are the color $SU(3)$ index. There are only two neutrino particle states -- since particles are defined to be invariant under the reversal operation and $R(\overline{\nu}_R) = P(\nu_L)$, a new particle does not result.}
\end{table}

To construct other particle families in a manner similar to that shown above, one needs to find other braids in $B_3$ such that acting with $C, P$ and $CP$ gives three other distinct braids, i.e. with no repetitions. It is obvious that none of the single crossing braids $\sigma_1, \sigma^{-1} _1, \sigma_2$ or $\sigma^{-1}_2$ are acceptable, nor is the trivial identity braid $\mathbb{1}$, since all are invariant under the parity operator $P$. Starting with $\sigma_1 \sigma_2$, the three operators result in the other two-crossing braids, $\sigma_2 \sigma_1, (\sigma_1 \sigma_2)^{-1}$ and $(\sigma_2 \sigma_1)^{-1}$. Thus, there is the possibility of another particle generation based on these four braid structures. For the braids with three crossings, the braid relation $\sigma_1 \sigma_2 \sigma_1 = \sigma_2 \sigma_1 \sigma_2$ (which is easily seen as true by drawing the respective braid pictures) cuts down the number of independent choices. However, this gives many possible choices for a starting braid, even at the level of only three crossings, such as $\sigma^2 _1 \sigma_2, \sigma^2 _1 \sigma^{-1} _2, \sigma^2 _1 \sigma_2$, and $\sigma ^2 _1 \sigma^{-1} _2$, resulting in a large number of particle generations. If current observations are taken seriously as a restriction on what the model should produce, this large number of possible families can be truncated to three total by looking only at the quotient braid group $B_3 (3)$. This is the group of 24 elements obtained by imposing the additional relationship $\sigma^3 _1 = \mathbb{1}$ (which implies $\sigma^3 _2 = \mathbb{1}$ as well), so that the only extra possibility for a generation from three-crossing braids is based on $\sigma^{-1} _1 \sigma_2 \sigma_1$ and its $C$ and $P$ partners. Although it is presented here without justification, one may speculate that imposing this extra relation on the braids may be connected to the reason the charge twists only appear in certain combinations.

As a final note, it is necessary to comment how altering the underlying braid structure of a particle generation affects the corresponding raising operators for their representations. Table \ref{alpha-def} shows the raising operators for the first generation of Standard Model particles; $\alpha^1$ and $\alpha^2$ act to permute the charge structure of a given particle, so they remain the same for all generations. However, $\alpha^3$ and $\alpha^4$, because they act to change one braid structure into another, are affected by the different underlying braids of each particle family. Although this would be a matter to determine in a final dynamical theory, it is likely that the mass of each gauge boson corresponding to the raising and lowering operators is somewhat dependent on topological factors, such as the number of crossings. Thus, this difference in the operators associated to $\alpha^3$ and $\alpha^4$ leads to the conclusion that some of the force-carrying particles, at least, may have different masses, depending on what generation they interact with. On the flip side, however, this may be a possible mechanism to explain issues such as the Cabibbo-Kobayashi-Maskawa matrix -- if the ``true'' physical states representing, for example, the $W^+$ boson, are a linear combination of quantum states realized by distinct braid structures (one for each particle generation), there may be a natural mixing of quark states from different generations, as well as an avenue to include photon and $Z$ boson states.

\section{Discussion}
\label{conclusions}

Below are comments on particular aspects of this model:

\begin{itemize}

	\item {\it Spin statistics:} As pointed out in previous work~\cite{Lambda}, there is no {\it a priori} way of dividing the particles given here into fermions and bosons. However, there is one interesting way of dividing them into these two types, based on the results given in this paper. Recall that for consistency, it was necessary for the raising operators to be invariant under parity; this cannot be true for the matter braids in order to get the entire family of particles. Thus it is interesting to note the division of the braid realization of the particles into those invariant under parity (corresponding to the fermions) and those that are not (giving the bosons). For example, compare the list of the force carrier particles given in Table \ref{alpha-def}, where the braid structure are invariant under a reflection of the crossing order, to the matter particles in Table \ref{SM-part}, where such a reflection takes a particle to one with opposite helicity.
	
	\item {\it CP invariance:} As a whole, the scheme presented here has the charge conjugation and parity operators commuting, so there is $CP$ invariance in that sense. In fact, using solely this, there would in fact be four neutrino states -- $\sigma_1 \sigma^{-1} _2 [0, 0, 0], \sigma_2 \sigma^{-1} _1 [0, 0, 0], \sigma^{-1} _1 \sigma_2 [0, 0, 0]$, and $\sigma^{-1} _2 \sigma_1 [0, 0, 0]$. Only the additional physical requirement of symmetry under the reversal operator $R$ eliminates the last two of these states as being distinct from the first two, and breaks $CP$ invariance.

	\item {\it Group structure:} The form of the braids given here, both those denoted as matter and gauge particles, in large part mimics the structure associated with the $SU(5)$ Lie algebra, i.e. particular representations, as well as the raising operators acting on them. This latter aspect can be pursued even further, since the raising operators should fit into the adjoint representation \rep{24} of $SU(5)$. Indeed, this is easily shown by writing each of the members of the \rep{24} as a sum of simple roots, and taking the corresponding product of braid realizations; the results are given in Table \ref{adjoint}. Note how each group of particles-- gluons, $X$ and $Y$ bosons -- have the same braid structure, with the charge structure going through all possible permutations; the ``missing'' permutations of $\{+1, 0, -1 \}$ are the other three gluons $g^2 _1, g^3 _1$, and $g^3 _2$, corresponding to negative weights in the adjoint \rep{24} representation.
	
	However, this is not a complete characterization of the $SU(5)$ Lie algebra, in that a definition for a commutation bracket is missing, as well as realizations for the elements of the Cartan subalgebra. The latter issue means that it is not clear whether this Lie group (if it can be rigorously defined) is $U(5)$, with one additional member that commutes with all the others, rather than $SU(5)$, or even a large group containing $SU(5)$. In this same vein, the action of the raising operators have been ``stopped by hand'' -- it is certainly possible to keep acting with any of the four $\alpha^i$ and get a new braid in $B_3$, just one outside the assumptions for this model. The question of why the raising operators annihilate certain particle realizations is something to be resolved by a proper dynamical theory.
	
	\item {\it Extra gauge bosons:} The minimal $SU(5)$ GUT features additional gauge bosons, the leptoquark $X$ and $Y$ particles, leading to the non-conservation of baryon and lepton number, although the difference in these, $B - L$, is conserved. This is seen in the particle associated with the $\alpha^3$ raising operator, which has a charge of $-4e/3$, and can convert leptons into quarks and vice versa. So, on the face of it, this gives rise to the same problem as the original GUT, namely the lack of experimental evidence for proton decay mediated by these types of particles. However, a potential saving grace is the following. Recall that the original Bilson-Thompson model imagined charges as twists on the strands of the braids; in order to have only zero or $\pm e/3$ charge on each braid, it was assumed that twists were taken modulo a $2 \pi$ rotation on the strand. In large part, this restriction is obeyed by the particles here, except for those with the same form for the $\alpha^3$ operator (with permutations of the charge structure). It is possible that this difference in the charge structure -- and subsequent lack of observational evidence for these particles -- may be due to the extreme improbability of the requisite topological moves to create such particle structures. Unfortunately, this intuition fails for the $Y$ bosons, which as shown in Table \ref{adjoint} have a trivial braid structure, so this matter is unresolved as well.

	\item {\it SO(10)}: The discussion in this paper used $SU(5)$ as the Lie group for the representations given, as the simplest choice beyond the Standard model group. However, it may be more natural to consider $SO(10)$ instead, where all the particles of a given handedness are in the same representation. The advantages of this would be the following. One of the potential issues with the $SO(10)$ GUT is the presence of an extra (right-handed) neutrino; this particle already appears in the braid formulation as the left-handed neutrino under the reversal symmetry $R$, so there is no need for an extra particle that is currently unobserved. In addition, for the $SU(5)$ scheme, in order to fit the relevant particles into the \rep{10} and $\repbar{10}$, the $\alpha^4$ operator is needed to take the left-handed down quark states to the left-handed up quarks, and the reverse for their anti-matter partners; these operators are not present for the \rep{5} and $\repbar{5}$ representations. However, since the identical raising operators $\alpha^i$ were shown to act on all four representations, it should be possible to use the $\alpha^4$ operator and take, e.g. the right-handed down quarks of \rep{5} to the right-handed up quarks of $\repbar{10}$. If this is not prevented for some unknown dynamical reason, this means these two representations naturally fit together, which leads to placing them together into the spinorial \rep{16} representation of $SO(10)$. This would mean the same braid realization would represent two of the simple roots of $SO(10)$, namely $\alpha^4$ and $\alpha^5$, which could be problematic, although it does imply there is no increase in the total number of gauge bosons.
 	
 	Finally, placing the Standard Model particles into the \rep{16} and $\repbar{16}$ representations for the $SO(10)$ GUT leads to a natural association with a direct product of five $SU(2)$ doublets~\cite{Wil-Zee82}. This fact arises because \rep{16} and $\repbar{16}$ are both spinor representations. It has been shown that the braid group $B_3$ has representations acting on $SU(2)$~\cite{Kau-Lom10}, so it is not unlikely, based on the results of this paper, that some version of the converse is true as well.
  
\end{itemize}

\begin{table}[hbt]
\begin{tabular}{cccc}
Particle 	& Structure								& Particle			& Structure									\\
\hline
$g^1 _2$		& $\mathbb{1} [+1, 0, -1]$				& $\overline{X}^3$		& $\sigma_2 \sigma^{-2} _1 \sigma_2 [-1, -2, -1]$			\\
$g^1 _3$		& $\mathbb{1} [0, +1, -1]$				& $\overline{Y}^1$		& $\mathbb{1} [0, 0, -1]$							\\
$g^2 _3$		& $\mathbb{1} [-1, +1, 0]$				& $\overline{Y}^2$		& $\mathbb{1} [-1, 0, 0]$							\\
$\overline{X}^1$	& $\sigma_2 \sigma^{-2} _1 \sigma_2 [-1, -1, -2]$	& $\overline{Y}^3$		& $\mathbb{1} [0, -1, 0]$							\\
$\overline{X}^2$	& $\sigma_2 \sigma^{-2} _1 \sigma_2 [-2, -1, -1]$	& $W^+$			& $\sigma^{-1} _2 \sigma^2 _1 \sigma^{-1} _2 [+1, +1, +1]$	\\
\end{tabular}
\caption{\label{adjoint}$SU(5)$ gauge bosons and structures for all positive weights in the adjoint representation \rep{24} of $SU(5)$; the negative roots correspond to the inverses of these. These states only act as raising operators for the first generation of particles; those with non-trivial braid structures would differ for other particle generations. Indices for particle names refer to their color $SU(3)$ index.}
\end{table}

In conclusion, this paper has shown that the first generation of Standard Model elementary particles, realized in the Bilson-Thompson model as a set of braids with three strands with charges as labels on these strands, consistently fit into the direct sum of representations $\rep{5} \oplus \repbar{5} \oplus \rep{10} \oplus \repbar{10}$ of the $SU(5)$ GUT. By using a particular definition of the parity operator $P$, along with charge conjugation $C$ obtained by taking a braid to its inverse, operators associated with the simple roots of $SU(5)$ were derived as braids themselves, leading to their interpretation as gauge bosons in this scheme. Only parity operators where the roots of the Lie algebra correspond to parity-invariant braids give a consistent picture, meaning there is a non-trivial structure developed here. These operators act on opposite sides of two representations that are conjugate to one other. From this, possible realizations of the other two particle families were given, and these would be the only additional possibilities if the braids are restricted to members of the quotient braid group $B_3 (3)$.

Although it is surprising how well this scheme works, it is still not a coherent model, since there is not a complete theory of the dynamics for these braids. Thus, many of the restrictions necessary to get only the Standard Model particles -- limits on the charges placed on the strands, or the breaking of $SU(5)$ into the Standard model symmetry group, for example -- are unexplained at this point, and must be left to conjectures on the actual realization used (if not the one here), or the action of unknown rules for topological changes in the particles. It is possible this could be done in the context of loop quantum gravity or another theory based on topological objects. A great deal of work is now needed to develop these ideas further into a rigorous theory testable with experimental data.

\end{document}